\documentclass[]{interact}

\usepackage{epstopdf}
\usepackage[caption=false]{subfig}

\theoremstyle{plain}

\theoremstyle{definition}

\theoremstyle{remark}
\newtheorem{remark}{Remark}

\newcommand{\Enc}{\mathsf{Enc}}
\newcommand{\Dec}{\mathsf{Dec}}

\usepackage{amsmath,amsfonts}
\usepackage{array}
\usepackage{textcomp}
\usepackage{stfloats}
\usepackage{url}
\usepackage{verbatim}
\usepackage{cite}
\usepackage{siunitx}
\usepackage{mathrsfs}
\usepackage{mathtools}
\mathtoolsset{showonlyrefs=true}
\usepackage{bm}
\usepackage{fancybox}
\usepackage{algpseudocode}
\usepackage{multirow}

\begin{document}

\fbox{
This work has been submitted to Advanced Robotics for possible publication.
}

\newpage

\articletype{FULL PAPER}

\title{Secure Motion-Copying via Homomorphic Encryption}

\author{
\name{Haruki Takanashi\textsuperscript{a}\thanks{CONTACT Haruki Takanashi. Email: takanashi@uec.ac.jp}, Kaoru Teranishi\textsuperscript{a, b}, and Kiminao Kogiso\textsuperscript{a}}
\affil{\textsuperscript{a}The Department of Mechanical and Intelligent Systems Engineering, The University of Electro-Communications, Chofu, Tokyo, Japan, \textsuperscript{b}Research Fellow of Japan Society for the Promotion of Science}
}

\maketitle

\begin{abstract}
This study aims to develop an encrypted motion-copying system using homomorphic encryption for secure motion preservation and reproduction. 
A novel concept of encrypted motion-copying systems is introduced, realizing the preservation, edition, and reproduction of the motion over encrypted data.
The developed motion-copying system uses the conventional encrypted four-channel bilateral control system with robotic arms to save the leader’s motion by a human operator in the ciphertext in a memory. 
The follower’s control system reproduces the motion using the encrypted data loaded from the secure memory.
Additionally, the developed system enables us to directly edit the motion data preserved in the memory without decryption using homomorphic operation. 
Finally, this study demonstrates the effectiveness of the developed encrypted motion-copying system in free motion, object contact, and spatial scaling scenarios.
\end{abstract}


\section{Introduction}
\textit{Motion-copying systems} record and reproduce human operations, leveraging bilateral control to extract motion data~\cite{yokokura2008,yukiyokokura2008}. 
The motion-copying systems find applications in fields such as tasks by skilled workers (e.g., spinning), calligraphy, and education~\cite{yokokura2008, yokokura2012, matsui2013a, matsui2013b, onoyama2014a}. 
Unlike motion capture and robot teaching, which rely solely on position data, the motion-copying system stores both position and force data. 
This duality facilitates the replication of intricate movements, notably in skilled labor and surgical procedures.

Numerous studies have been dedicated to enhancing the motion-copying system. 
For example, spatial scaling (resizing motion trajectories) and time scaling (variable playback speeds) have been studied extensively in~ \cite{yukiyokokura2008, kuwahara2010, miura2014, igarashi2014, igarashi2015, okano2018}. 
In~\cite{hiroyukitanaka2009, nakano2015}, the authors have explored the methods of reducing the size of stored motion data and have developed efficient data compression techniques. 
Furthermore, there have been studies on modeling the environment via the motion data when the follower robot contacts with the environment~\cite{yokokura2009a, kobayashi2022}. 
In~\cite{yokokura2010a, ohnishi2011}, the authors have investigated methods for efficiently searching for motions that match the current environment and configuration  within an accumulated database. 
In~\cite{matsui2013b, phuong2014a}, the motion-copying system between robots with different structures has been developed.
Additionally, the stability of motion-copying systems is analyzed in~\cite{yokokura2009}.

However, the network involved in the bilateral control system is needed for operating a motion-copying system effectively.
Concerns regarding the security of such networked systems arise  when there are malicious attackers on the network. 
There have been instances of industrial control systems being targeted in cyberattacks, such as Stuxnet and Industroyer~\cite{Stuxnet2011, Industroyer2016}. 
As a countermeasure to cyberattacks on control systems, an \textit{encrypted control} method that operates on encrypted data using homomorphic encryption~\cite{elgamal1985,paillier1999, emura2018} has been proposed in~\cite{Kogiso2015, schlze2021en}. 
Encrypted control involves encrypting input and output signals and parameters using a specific homomorphic encryption scheme, providing effective protection against eavesdropping and tampering attacks. 
There have been several application studies on encrypted control, such as PID control systems~\cite{Cheon2018, miyamoto2023} and encrypted bilateral control systems~\cite{Takanashi2023, shono2022}, to enhance cybersecurity for remote control and operation.
Enhancing the cybersecurity of motion-copying systems by applying homomorphic encryption is crucial, especially when dealing with sensitive and confidential data such as the movements of skilled workers.

\begin{figure}[t!]
\centering
\includegraphics[width=0.65\linewidth]{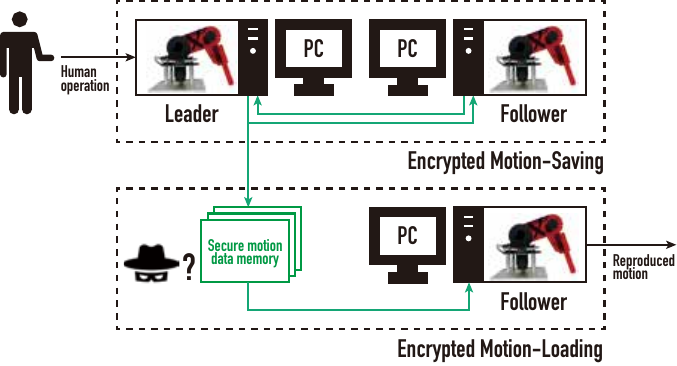}
\caption{Conceptual diagram of the encrypted motion-copying system with robotic arms to securely preserve and reproduce human operation, where green lines and blocks indicate encrypted communications and processes.}
\label{fig:concept}
\end{figure}

The objective of this study is to develop an encrypted motion-copying system using homomorphic encryption for secure motion preservation and reproduction. 
We introduce a novel concept of encrypted motion-copying systems, presented in Fig.~\ref{fig:concept}, where encrypted signals and control parameters are highlighted in green.
The concept is an extension of the conventional methods~\cite{yokokura2008,matsui2013} that realizes storing, editing, and reproducing the motion while encrypted.
Based on this framework, this study develops the encrypted motion-copying system using the encrypted four-channel bilateral control system with the leader and follower robotic arms from \cite{tanaka2023}.
The encrypted motion-copying system comprises the motion-saving and motion-loading phases. 
In saving motion, the system uses the encrypted bilateral control system to save the leader’s motion by a human operator in the secure memory in the ciphertext. 
In loading motion, the follower’s control system reproduces the saved motion.
Additionally, the proposed system allows for editing the encrypted stored motion data directly without decryption, facilitating the generation of appropriate motions. 
By maintaining motion in ciphertext, eavesdropping attempts by malicious entities are thwarted.
Finally, this study demonstrates the effectiveness of the developed encrypted motion-copying system in free motion, object contact, and spatial scaling scenarios.

The contributions of this paper are as follows.
i) This study presents the new concept of an encrypted motion-copying system, which can be realized using the conventional encrypted four-channel bilateral control system.
The developed system is the first application of the encrypted control fashion to motion-copying.
ii) This study reveals that homomorphic encryption helps establish a secure framework of motion-copying, bolstering cybersecurity for motion preservation and reproduction.

The reminder structure of this paper is as follows.
Section~\ref{sec:pre} provides preliminary information about the encrypted bilateral control system.
Section~\ref{sec:encmc} proposes the encrypted motion-copying system involving the saving and loading phases and how to scale the motion data.
Section~\ref{sec:er} demonstrates the developed encrypted motion-copying system to confirm the motion reproduction.
Finally, Section~\ref{sec:con} concludes this paper.

\section{Preliminaries}\label{sec:pre}
This section provides preliminary information about the homomorphic encryption scheme proposed in~\cite{elgamal1985}, the maps between real and integer numbers to encode and decode used in~\cite{Kogiso2015}, and the encrypted bilateral control system developed in~\cite{Takanashi2023}.

\subsection{Multiplicatively homomorphic encryption}
The ElGamal encryption scheme, denoted as $\mathcal{E}=(\mathsf{Gen},\mathsf{Enc},\mathsf{Dec})$, consists of three fundamental algorithms: key generation $\mathsf{Gen}$, encryption $\mathsf{Enc}$, and decryption $\mathsf{Dec}$.
With a key length of $\lambda$ bits, the key generation algorithm, $\mathsf{Gen}$, produces a public key $k_p$ and a private key $k_s$, which can be expressed as $\mathsf{Gen}(\lambda) = (k_p, k_s)$.
The encryption algorithm, $\mathsf{Enc}$, is responsible for encrypting a plaintext $m$ using the public key $k_p$, resulting in ciphertext $c$, i.e., $\mathsf{Enc}(k_p, m) = c$.
The decryption algorithm, $\mathsf{Dec}$, is employed to decrypt the ciphertext $c$ using both the public key $k_p$ and the private key $k_s$, and it yields the plaintext $m'$, i.e., $\mathsf{Dec}(k_s, c) = m'$. 
It's important to note that the equation $\mathsf{Dec}(k_s, \mathsf{Enc}(k_p, m)) = m$ holds true.

ElGamal encryption possesses a multiplicative homomorphic property, allowing for ciphertext multiplication. 
Given two plaintexts, $m_1$ and $m_2$, and their respective ciphertexts, $c_1 = \mathsf{Enc}(k_p, m_1)$ and $c_2 = \mathsf{Enc}(k_p, m_2)$, the resulting ciphertext of their product $m_1 m_2$ is expressed as $\mathsf{Enc}(k_p, m_1 m_2) = c_1 \otimes c_2$, where $\otimes$ represents element-wise multiplication. 
This property is known as a multiplicative homomorphism.

\subsection{Encoding and decoding}
To utilize ElGamal encryption, real numbers must be encoded to the plaintext space $\mathcal{M}$ using an appropriate quantization gain $\gamma$ and $a$, as defined by the following function: 
\begin{align}
\textrm{Encoder}: \mathbb{R} \ni x \mapsto \check{x} = \lceil \gamma x + a \rfloor \in \mathcal{M},
\end{align}
where $a=0$ if $x\geq 0$ and $a=p$ otherwise, and
$\lceil \cdot \rfloor$ signifies rounding to the nearest element in the plaintext space $\mathcal{M}$.
If $q$ and $p = 2q + 1$ are prime, $q$ and $p$ are called a Sophie Germain prime and a safe prime number, respectively. 
Such primes play a crucial role in enhancing the security of ElGamal encryption.
Furthermore, the decrypted plaintext is decoded as follows:
\begin{align}
\textrm{Decoder}: \mathcal{M} \ni \check{x} \mapsto \tilde{x} = \frac{\check{x} - b}{\gamma} \in \mathbb{Q},
\end{align}
where $b = 0$ if $\check{x} \leq q$ and $a = p$ otherwise and $\mathbb{Q}$ denotes rational numbers. 
Hereafter, we redefine $\Enc$ to perform encoding and encryption and $\Dec$ to perform decryption and decoding.

\begin{figure}[tb]
  \centering
\includegraphics[scale=.65]{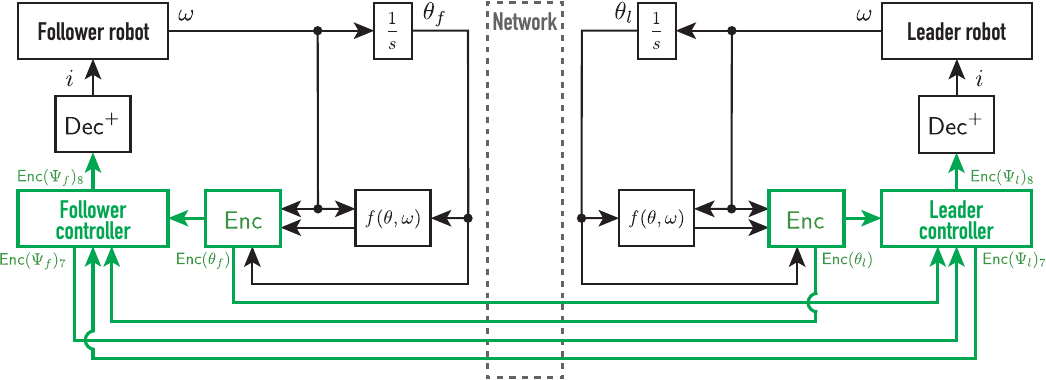}
  \caption{Block diagram of encrypted four-channel bilateral control system~\cite{Takanashi2023}, where the green blocks represent homomorphic operations.}
  \label{fig:bd_4ch}
\end{figure}

\subsection{Encrypted bilateral control system}\label{sec:enc4ch}
The encrypted four-channel bilateral control system with robotic arms, as illustrated in Fig.~\ref{fig:bd_4ch}, is used to construct the proposed secure motion-copying system. 
The control objective is to achieve ideal kinesthetic coupling:
\begin{align}
&\theta_l(t) - \theta_f(t) = 0,\quad \tau^{e}_l(t) + \tau^{e}_f(t) = 0, \label{math5}
\end{align}
between the yaw-axis motors of the leader and follower robotic arms:
\begin{equation}
\bar{J}\ddot{\theta_l}(t) = \bar{K} i_l(t) - \tau_l^{d}(t),\quad 
\bar{J}\ddot{\theta_f}(t) = \bar{K} i_f(t) - \tau_f^{d}(t), 
\label{math:plant} 
\end{equation}
where $t\geq 0$ is a continuous time, 
$\theta\in\mathbb{R}$ is the angle measured by the encoder; 
$\tau^{d}\in\mathbb{R}$ represents the unknown disturbance torque;
$i\in\mathbb{R}$ represents the current input to the motor;
$\bar{J}$ and $\bar{K}$ represent the nominal moment of inertia and torque coefficient of the motor, which are described in its specifications;
subscripts $l$ and $f$ denote the leader and follower, respectively.
The torque is expressed in the following equation consisting of modeling error, cable tension, friction, and an external force $\tau^e$: 
\begin{align}
\tau^{d}_l(t) &:= (J - \bar{J}) \ddot{\theta}_l(t) + (\bar{K} - K) i_l(t) + f(\theta_l(t), \omega_l(t)) + \tau^{e}_l(t),\\
\tau^{d}_f(t) &:= (J - \bar{J}) \ddot{\theta}_f(t) + (\bar{K} - K) i_f(t) + f(\theta_f(t), \omega_f(t)) + \tau^{e}_f(t),
\end{align}
where $J$ and $K$ represent the moment of inertia and the torque coefficient of the motor, which are not measurable, respectively, and
a nonlinear function $f$ of the angle $\theta$ and velocity $w=\dot\theta$, encompasses cable tension and friction forces:
\begin{align}
f(\theta_l(t), \omega_l(t)) &:= a_1 \theta_l(t) + a_2 \tan^{-1}{(a_3\omega_l(t) + a_4)} +a_5,\\
f(\theta_f(t), \omega_f(t)) &:= a_1 \theta_f(t)+ a_2 \tan^{-1}{(a_3\omega_f(t) + a_4)} + a_5.
\end{align}

To achieve the control objective for the robotic arm~\eqref{math:plant}, we designed model-based linear controllers consisting of Proportional-Derivative (PD) and Proportional (P) controllers for position and force control, Disturbance Observer (DOB), and Reaction Force Observer (RFOB).
Discretizing the controllers by a bilinear transformation with a sampling period $T_s$ results in the following discrete-time state-space representation:
\begin{align}
   \psi_{lk} = \Phi \xi_{lk} \coloneqq f_l(\Phi, \xi_{lk}), \quad \psi_{fk} = \Phi \xi_{fk} \coloneqq f_f(\Phi, \xi_{fk}),
   \label{math:ctrl}
\end{align}
with 
\begin{align}
  & \psi_{lk} \coloneqq
  \begin{bmatrix}
    x_{l(k+1)} \\
    u_{lk}
  \end{bmatrix} ,\ 
  \psi_{fk} \coloneqq
  \begin{bmatrix}
    x_{f(k+1)} \\
    u_{fk}
  \end{bmatrix},\ 
  \xi_{lk} \coloneqq
  \begin{bmatrix}
    x_{lk} \\
    v_{lk}
  \end{bmatrix},\ 
  \xi_{fk} \coloneqq
  \begin{bmatrix}
    x_{fk} \\
    v_{fk}
  \end{bmatrix},\
  \Phi =
  \begin{bmatrix}
    A_c & B_c \\
    C_c & D_c
  \end{bmatrix}, \label{math:xil} 
  \end{align}
  \begin{align}
  &  x_{l} =
  \begin{bmatrix}
    e_{l} \\
    \dot{e}_{l} \\
    q_{l} \\
    z_{l} \\
    i_{l}
  \end{bmatrix},\,
     x_{f} =
  \begin{bmatrix}
    e_{f} \\
    \dot{e}_{f} \\
    q_{f} \\
    z_{f} \\
    i_{f}
  \end{bmatrix},\,
  u_{l} =
  \begin{bmatrix}
    \hat{\tau}^{d}_{l} \\
    \hat{\tau}^{e}_{l} \\
    i_{l}
  \end{bmatrix},\,
    u_{f} =
  \begin{bmatrix}
    \hat{\tau}^{d}_{f} \\
    \hat{\tau}^{e}_{f} \\
    i_{f}
  \end{bmatrix},\,
  v_l =
  \begin{bmatrix}
    \theta_f \\
    \theta_l \\
    \omega_l \\
    \hat{\tau}^{e}_f \\
    \hat{\tau}^{e}_l \\
    f(\theta_l, \omega_l)
  \end{bmatrix},\,
   v_f =
  \begin{bmatrix}
    \theta_l \\
    \theta_f \\
    \omega_f \\
    \hat{\tau}^{e}_l \\
    \hat{\tau}^{e}_f \\
    f(\theta_f, \omega_f)
  \end{bmatrix}, 
\end{align}
where subscript $k\geq 0$ represents a step (discrete time), $\psi_{l}\in\mathbb{R}^{8}$ and $\psi_{f}\in\mathbb{R}^{8}$ are controller outputs for the leader and follower, respectively;
$\xi_{l}\in\mathbb{R}^{11}$ and $\xi_{f}\in\mathbb{R}^{11}$ are the controller inputs for the leader and follower, respectively;
$\Phi\in\mathbb{R}^{8\times11}$ is identical controller parameters (system matrix) for the leader and follower.
Moreover, $f_l$ and $f_f$ indicate the leader and follower controllers, respectively. 

Let $f_l = f_l^{+} \circ f_l^{\times}$, where $f_l^{+}$ and $f_l^{\times}$ denote addition and multiplication on the leader, respectively, and similarly, $f_f=f_f^{+} \circ f_f^{\times}$ is defined on the follower. 
If the quantization error is negligibly small, the leader and follower encrypted controllers $f^{\times}_{l\mathcal{E}^+}$ and $f^{\times}_{f\mathcal{E}^+}$ are represented as follows:
\begin{align}
    f^{\times}_{l\mathcal{E}^+}: (\mathsf{Enc}({\Phi}),~\mathsf{Enc}({\xi}_{lk})) \mapsto \mathsf{Enc}({\Psi}_{lk}),\ \ 
    f^{\times}_{f\mathcal{E}^+}: (\mathsf{Enc}({\Phi}),~\mathsf{Enc}({\xi}_{fk})) \mapsto \mathsf{Enc}({\Psi}_{fk}),
    \label{math:encf}
\end{align}
where $\Psi_{lk} \coloneqq f_l^\times(\Phi, \xi_{lk})$ and $\Psi_{fk} \coloneqq f_l^\times(\Phi, \xi_{fk})$, which are computed using the multiplicative homomorphism.
Define $\mathsf{Dec}^+ \coloneqq f^+ \circ \mathsf{Dec}$. 
Then, $\tilde{\psi}_{lk}$ and $\tilde{\psi}_{fk}$ are obtained as follows:
\begin{align}
    \mathsf{Dec}^+(\mathsf{Enc}(\Psi_{lk})) &= {\tilde\psi}_{lk},\quad
    \mathsf{Dec}^+(\mathsf{Enc}(\Psi_{fk})) = {\tilde\psi}_{fk}.\label{math:decplus}
\end{align}
where $\mathcal{E}^+ \coloneqq (\mathsf{Gen}, \mathsf{Enc}, \mathsf{Dec}^+)$ is a modified ElGamal encryption scheme.
In addition, the quantization error, denoted as $\psi_l-{\tilde\psi}_l$, is assumed to be negligible in this study, and the definition and derivation of the matrices and vectors are detailed in~\cite{Takanashi2023}.
Consequently, the encrypted four-channel bilateral control system with the robotic arms, encrypted controller \eqref{math:encf}, and modified decryption \eqref{math:decplus}, can be constructed as illustrated in Fig.~\ref{fig:bd_4ch}, where it features encrypted communications and blocks, highlighted in green.

\section{Encrypted Motion-Copying System}\label{sec:encmc}
\begin{figure}[tb]
  \centering
  \subfloat[Encrypted motion-saving system]{\includegraphics[width=.7\linewidth]{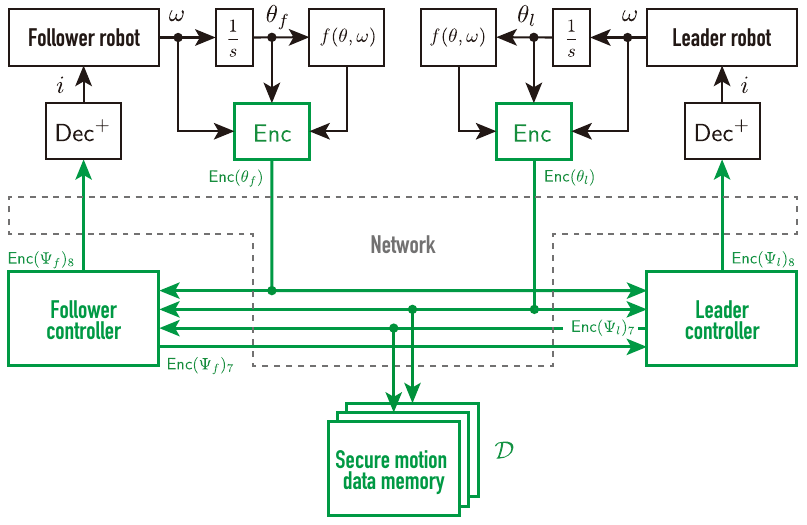}%
  \label{fig:bd_emss}}\\
  \subfloat[Encrypted motion-loading system]{\includegraphics[width=.7\linewidth]{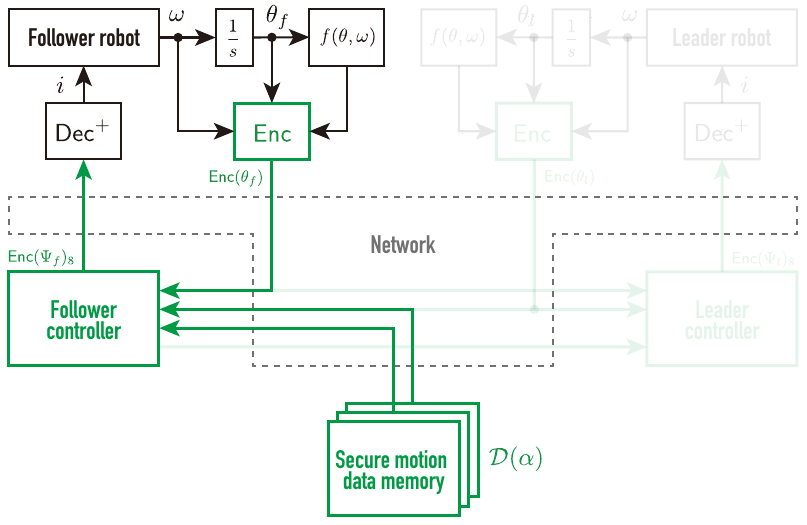}%
  \label{fig:bd_emls}}
  \caption{Block diagrams of the proposed encrypted motion-copying system.}
  \label{fig:bd_emcs}
\end{figure}

This section proposes a concept of encrypted motion-copying systems capable of storing motion data in a ciphertext and editing the stored data without decryption to reproduce the motion.
The proposed motion-copying system consists of three phases: saving, spatial scaling, and loading the motion.
The saving and loading phases consider the motion-saving and motion-loading systems illustrated in Fig.~\ref{fig:bd_emcs}.
To realize the proposed concept, this study uses the encrypted four-channel bilateral control system described in \textbf{Section~\ref{sec:enc4ch}}.

\subsection{Secure motion data memory}
Firstly, the secure motion data memory is introduced to store and edit the motion data.
The memory is supposed to be located on a specific networked server that the motion-saving and motion-loading systems can communicate with, and the secure motion data on the memory is denoted as a finite set of data sequences:
\begin{align}
    \mathcal{D} = \left\{\{x_k\}_{k=0}^N,\{y_k\}_{k=0}^N\right\},
\end{align}
where $x_k$ and $y_k$ are a ciphertext at step $k$ and $\{x_k\}_{k=0}^N$ means a sequence of encrypted signal $x_k$ with data length $N$.
The multiplicative homomorphism of the ElGamal encryption scheme allows us to scale the data set by scaling parameter $\alpha=(\alpha_x,\alpha_y)\in\Re^2$ with the corresponding quantization gain $\gamma_\alpha>0$.
This operation is referred to as spatial scaling~\cite{miura2014}.
When the motion data are scaled, the updated data set is denoted as 
\begin{align}
\mathcal{D}(\alpha)=\left\{\{x_k\otimes\Enc(\alpha_x)\}_{k=0}^N,\{y_k\otimes\Enc(\alpha_y)\}_{k=0}^N\right\}.
\end{align}
It should be noted that the motion data, $\mathcal{D}$ and $\mathcal{D}(\alpha)$, and scaling parameter, $\alpha$, are all encrypted and stored in the memory.
In addition, although the proposed concept enables us to store multiple motion-data sequences with different data lengths in the memory, this paper considers two sequences with identical lengths for simplicity.

\begin{remark}
Spatial scaling is required when the size of the follower robot differs between motion-saving and motion-loading phases. 
This study achieves spatial scaling using multiplicative homomorphic encryption. 
Meanwhile, the temporal scaling that adjusts timestamps regarding the motion data, proposed~\cite{igarashi2014}, involves addition and multiplication, so a fully homomorphic encryption scheme is required to realize our proposed concept. 
This extension will be future work to address editing the motion data for several conditions.
\end{remark}

\subsection{Encrypted motion-saving system}
We present the encrypted motion-saving system, illustrated in Fig.~\ref{fig:bd_emcs}(a), consisting of the encrypted four-channel bilateral control system and the secure motion data memory to save the leader’s motion data.
This study considers that the motion data measured during the operation are the encrypted rotation angle $\Enc(\theta_{lk})$ and estimated torque $\Enc(\hat{\tau}^e_{lk})$ regarding the yaw-axis motor, which appears in the argument and output of the leader’s encrypted controller $f^{\times}_{l\mathcal{E}^+}$ in \eqref{math:encf}.
They are encapsulated within the set denoted as $\mathcal{D}$:
\begin{align}
    \mathcal{D} = \left\{\{\Enc{(\theta_{lk})\}_{k=0}^N}, \{\Enc{(\hat{\tau}^e_{lk})\}_{k=0}^N}\right\}.
    \label{D}
\end{align}

Fig.~\ref{fig:bd_emcs}(a) confirms that the presented motion-saving system is a secure networked system configuration.
The green lines and blocks indicate the encrypted communications and data processing, so the controllers and memory outside the plant sides operate with encrypted signals and parameters.
Hence, even if malicious attackers eavesdrop on the control system or steal data from the secure motion data memory, it is challenging to ascertain the stored motions.

\subsection{Encrypted motion-loading system}
We present the encrypted motion-loading system, illustrated in Fig.~\ref{fig:bd_emcs}(b), consisting of the follower-side control system with the robotic arm and the secure motion data memory to reproduce the motion.
This study considers that the motion data $\mathcal{D}$ in \eqref{D} are computed using an appropriate scaling parameter to be updated, as follows:
\begin{align}
\mathcal{D}(\alpha)=\left\{\{\Enc(\theta_{lk})\otimes\Enc(\alpha_x)\}_{k=0}^N,\{\Enc(\hat{\tau}_{lk}^e)\otimes\Enc(\alpha_y)\}_{k=0}^N\right\}.
\end{align}
When the follower wants to reproduce the motion, the motion-loading system transmits the secure motion data of $\mathcal{D}(\alpha)$ from the memory to the follower’s controller.
Noted that the decoder installed on the follower side must have quantization gain set to $\gamma=\gamma_\xi\gamma_\Phi\gamma_\alpha$, where $\gamma_\xi>0$ and $\gamma_\Phi>0$ are quantization gains corresponding to encoding signals $\xi$ and parameters $\Phi$, respectively.
Then, the follower robotic arm is controlled by the decrypted control command regarding an input current, which reproduces the desired motion.

Fig.~\ref{fig:bd_emcs}(b) confirms that the presented motion-loading system is also a secure networked system configuration because the follower controller and memory outside the plant side are operated with the signals and parameters encrypted.
Hence, it is challenging for malicious attackers to ascertain the stored motions.

\begin{remark}
The introduced concept in this study can be applied to a different motion-loading situation where the follower’s controller is unencrypted and located on the plant side. 
In this setup, the decryption and encryption blocks on the plant side are not needed; 
The signal $\theta_f$ is input directly to the follower controller; 
The secure motion data loaded from the memory are decrypted in receiving on the follower controller, and the decrypted data are used to control processing.
This implementation implies that the follower control system is not necessarily encrypted from the viewpoint of secure motion-copying.
\end{remark}

Consequently, the concept of the proposed motion-copying system allows for the secure preservation and reproduction of motion. 
In the following section, we will delve into the practical setup used to investigate the effectiveness of the proposed system.

\section{Experimental validation}\label{sec:er}
This section demonstrates the effectiveness of the proposed encrypted motion-copying system through three scenarios: free motion, object contact, and spatial scaling. 
This study conducted experimental validation because it is challenging to model human-robot and robot-environment interactions, such as human movements and the response of the contact environment.
The experiments set the sampling period $T_s$, data length $N$, and key length $\lambda$ to $\SI{20}{ms}$, 750, and $\SI{128}{bit}$, taking computer resources and real-time computation into account.
The quantization gains $\gamma_\xi$, $\gamma_\Phi$, and $\gamma_\alpha$ were all set to $10^6$, in trial and error.
The controller parameters in $\Phi$ were as follows,
\begin{align}
A_c&=\begin{bmatrix} 0 & 0 & 0 & 0 & 0 \\ -50 & 0 & 0 & 0 & 0 \\ 0 & 0 & 0.333333 & 0 & 0.333333 \\ 0 & 0 & 0 & 0.333333 & 0.333333 \\ -0.615 & 0 & 1.333333 & 0 & 0.333333 \end{bmatrix},
\end{align}
\begin{align}
B_c&=\begin{bmatrix} 1 & -1 & 0 & 0 & 0 & 0 \\ 50 & -50 & 0 & 0 & 0 & 0 \\ 0 & 0 & 0.0013667 & 0 & 0 & 0 \\ 0 & 0 & 0.0013667 & 0 & 0 & -0.666667 \end{bmatrix},\\[1ex]
C_c&=\begin{bmatrix} 0 & 0 & 0.666667 & 0 & 0.1666667 \\ 0 & 0 & 0 & 0.666667 & 0.1666667 \\ -0.615 & 0 & 1.333333 & 0 & 0.333333 \end{bmatrix},\\[1ex]
D_c&=\begin{bmatrix} 0 & 0 & -0.001367 & 0 & 0 & 0 \\ 0 & 0 & -0.001367 & 0 & 0 & -0.333333 \\ 0 & 0 & 0.0013667 & 0 & 0 & -0.666667 \end{bmatrix}.
\end{align}
For PC controller specifications and other parameters, please refer to~\cite{Takanashi2023}.

\subsection{Free motion}
\begin{figure}[t!]
\centering
\subfloat[Rotation angle: $\theta_l$ and $\theta_f$.]{%
\resizebox*{.24\textwidth}{!}{\includegraphics{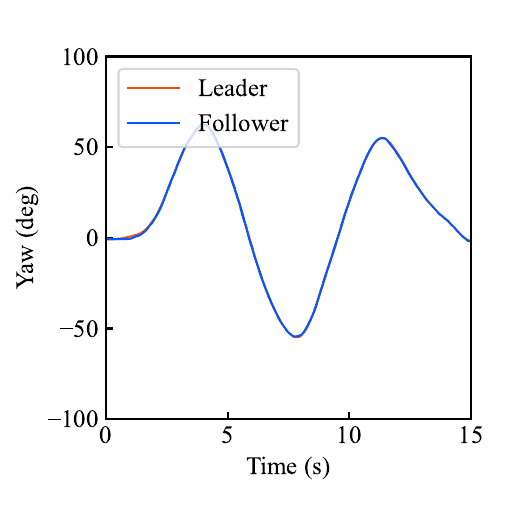}}}\  
\subfloat[Estimated reaction force: $\hat{\tau}^{e}_l$ and $\hat{\tau}^{e}_f$.]{%
\resizebox*{.24\textwidth}{!}{\includegraphics{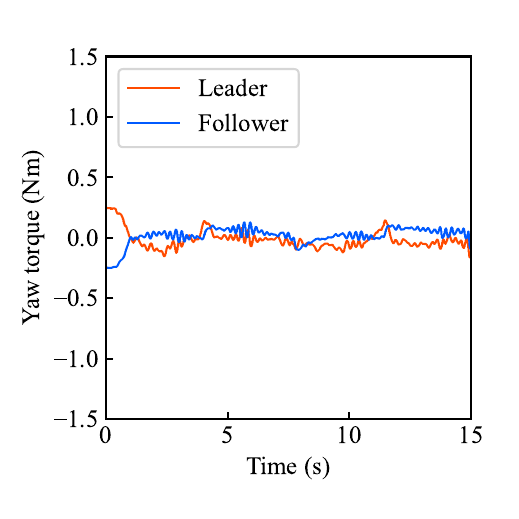}}}\ 
\subfloat[Encrypted rotation angle: $\Enc{(\theta_l)}$.]{%
\resizebox*{.24\textwidth}{!}{\includegraphics{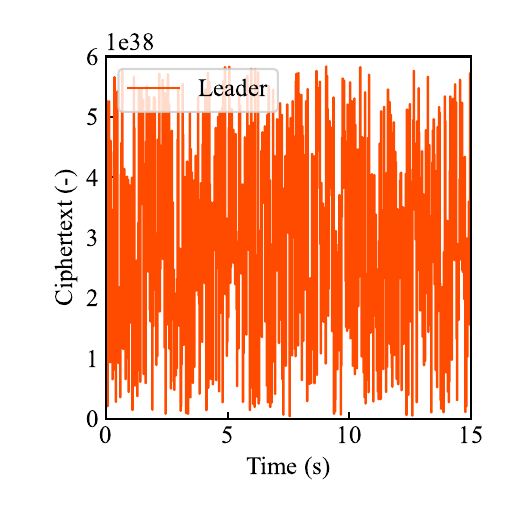}}}
\subfloat[Enrypted estimated reaction force: $\Enc{(\hat{\tau}^{e}_l)}$.]{%
\resizebox*{.24\textwidth}{!}{\includegraphics{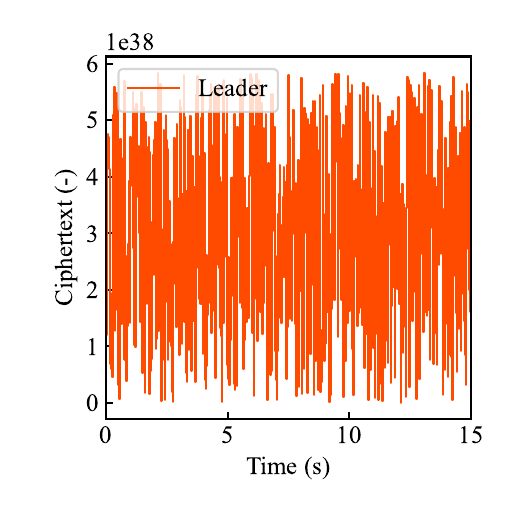}}}\\
\subfloat[Rotation angle: $\theta_l$ and $\theta_f$.]{%
\resizebox*{.24\textwidth}{!}{\includegraphics{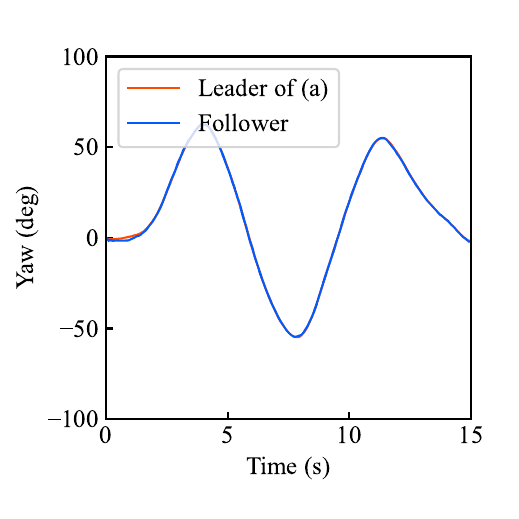}}}\ 
\subfloat[Estimated reaction force: $\hat{\tau}^{e}_l$ and $\hat{\tau}^{e}_f$.]{%
\resizebox*{.24\textwidth}{!}{\includegraphics{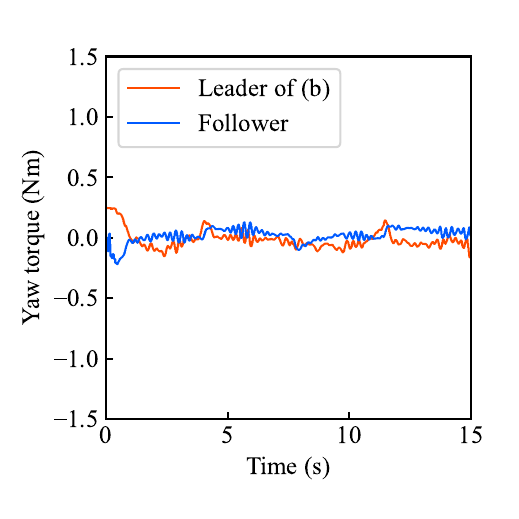}}}\ 
\subfloat[Encrypted rotation angle: $\Enc{(\theta_l)}$.]{%
\resizebox*{.24\textwidth}{!}{\includegraphics{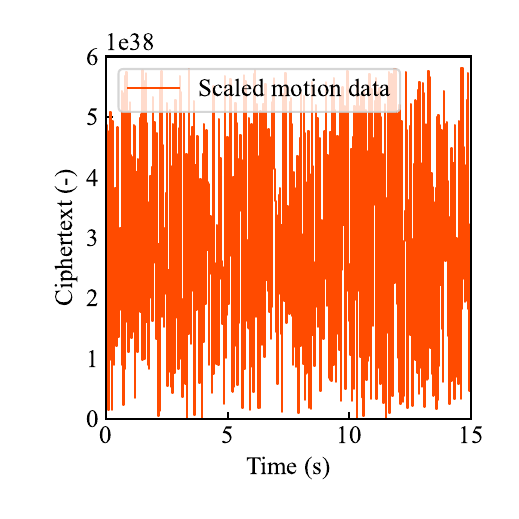}}}
\subfloat[Enrypted estimated reaction force: $\Enc{(\hat{\tau}^{e}_l)}$.]{%
\resizebox*{.24\textwidth}{!}{\includegraphics{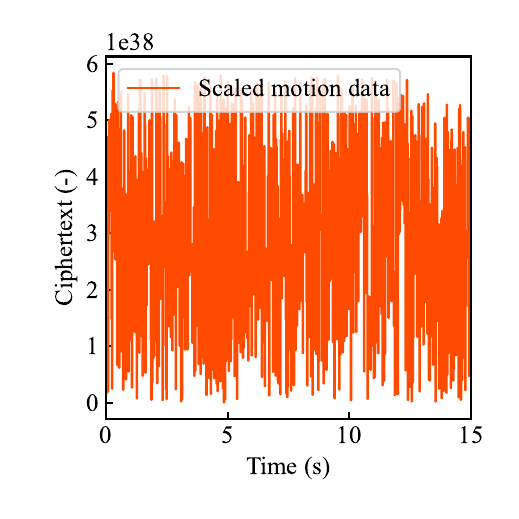}}}
\caption{The experimental results for free motion scenario; (a)-(d) saving motion; (e)-(h) loading motion} \label{pt}
\end{figure}

The free motion scenario represents saving and loading motion data without physical interaction with objects. 
In saving motion, a human operator controls the leader robot, and the follower robot replicates its movements, resulting in a 15-second motion sequence regarding the leader that is stored in the secure motion data memory. 
In loading motion, the developed system reproduces the stored motion such that the follower emulates the leader’s actions, where the scaling parameter $\alpha$ was set to (1,1).
The obtained results for saving and loading the motion are shown in Fig.~\ref{pt}.

Figs.~\ref{pt}(a)-(d) show the time responses of the leader and follower robotic arms during saving motion, where the red and blue lines indicate the leader and follower, respectively.
Figs.~\ref{pt}(a) and (b) illustrate the yaw-axis rotation angles and the estimated torque on the leader and follower sides, respectively.
Figs.~\ref{pt}(c) and (d) show the encrypted rotation angle and estimated torque of the leader, respectively.
The figures confirm that the follower’s behavior almost matches the leader’s and that the torques are approximately zeros due to no contact.
The encrypted motions regarding $\theta_l$ and $\hat{\tau}^{e}_l$ shown in Figs.~\ref{pt}(c) and (d) were stored in the secure motion data memory $\mathcal{D}$, where the encrypted data exhibit random fluctuations at the order of $10^{38}$.

Figs.~\ref{pt}(e)-(h) show the time responses of the follower during loading motion from $\mathcal{D}(\alpha)$. 
Figs.~\ref{pt}(e) and (f) illustrate the yaw-axis rotation angles and the estimated torques, respectively, where the blue line indicates the resulting follower’s behavior and the red line being the same as one of Figs.~\ref{pt}(a) and (b) was added for comparison.
Figs.~\ref{pt}(g) and (h) show the encrypted rotation angles and estimated torques that were loaded from $\mathcal{D}(\alpha)$, respectively.
Consequently, the figures confirm that the reproduced behavior of the follower has a good tracking performance to the original behavior and that the estimated torque of the follower is almost the same as that of the follower in saving motion.

\subsection{Object contact}
\begin{figure}[t!]
\centering
\includegraphics[width=0.4\linewidth]{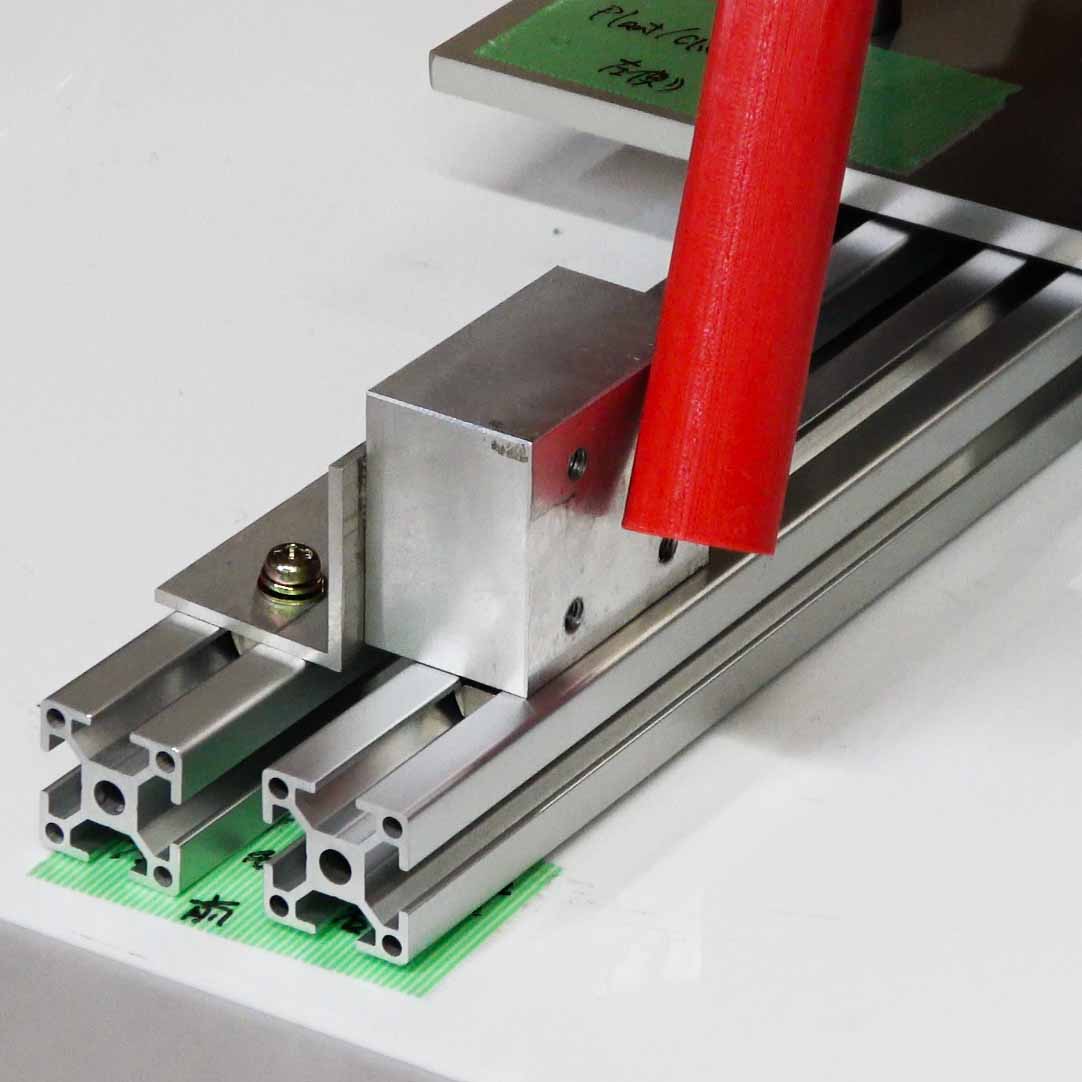}
\caption{The robotic arm making contact with an aluminum block~\cite{Takanashi2023}.}
\label{obst}
\end{figure}

\begin{figure}[t!]
\centering
\subfloat[Rotation angle: $\theta_l$ and $\theta_f$.]{%
\resizebox*{.24\linewidth}{!}{\includegraphics{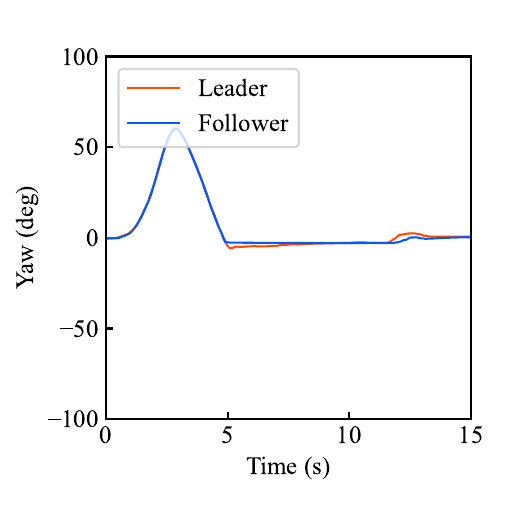}}}\ 
\subfloat[Estimated reaction force: $\hat{\tau}^{e}_l$ and $\hat{\tau}^{e}_f$.]{%
\resizebox*{.24\linewidth}{!}{\includegraphics{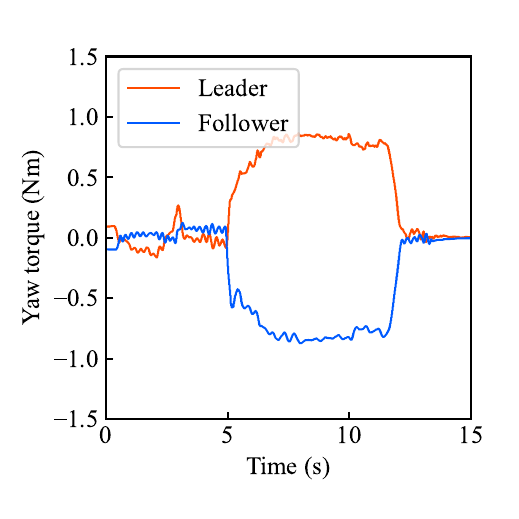}}}\ 
\subfloat[Encrypted rotation angle: $\Enc{(\theta_l)}$.]{%
\resizebox*{.24\linewidth}{!}{\includegraphics{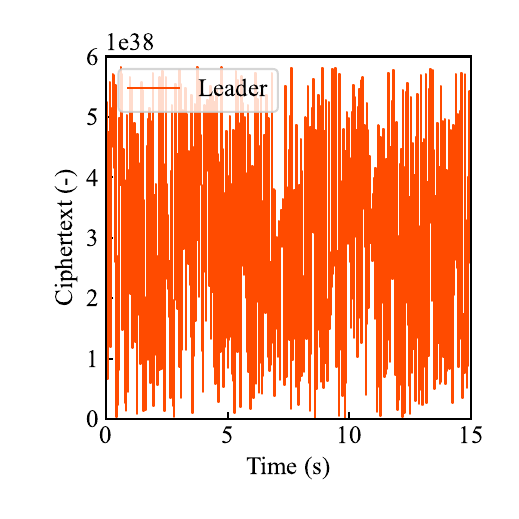}}}
\subfloat[Enrypted estimated reaction force: $\Enc{(\hat{\tau}^{e}_l)}$.]{%
\resizebox*{.24\linewidth}{!}{\includegraphics{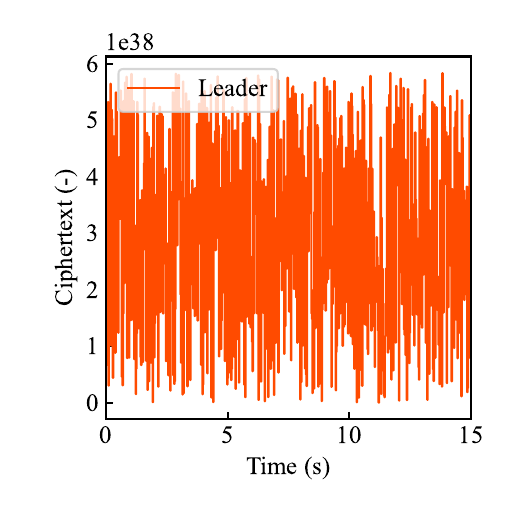}}}\\
\subfloat[Rotation angle: $\theta_l$ and $\theta_f$.]{%
\resizebox*{.24\linewidth}{!}{\includegraphics{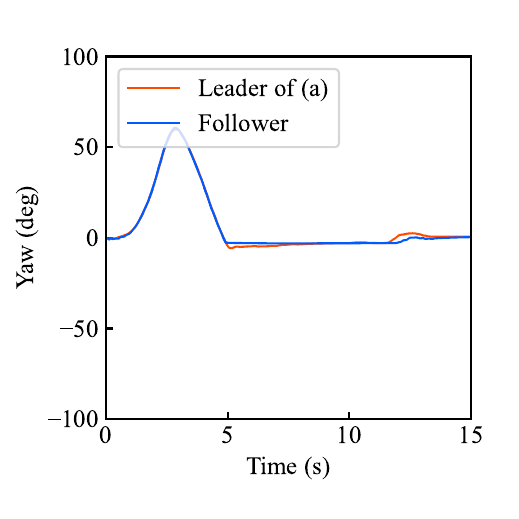}}}
\subfloat[Estimated reaction force: $\hat{\tau}^{e}_l$ and $\hat{\tau}^{e}_f$.]{%
\resizebox*{.24\linewidth}{!}{\includegraphics{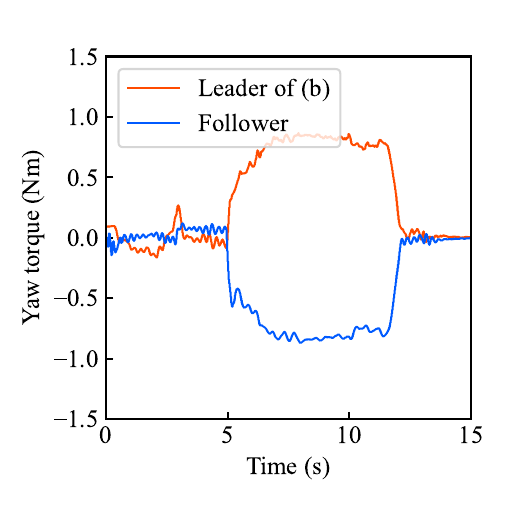}}} 
\subfloat[Enrypted rotation angle: $\Enc{(\theta_l)}$.]{%
\resizebox*{.24\linewidth}{!}{\includegraphics{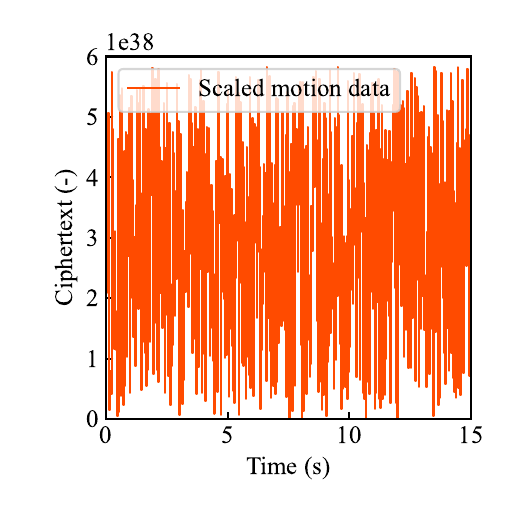}}}
\subfloat[Enrypted estimated reaction force: $\Enc{(\hat{\tau}^{e}_l)}$.]{%
\resizebox*{.24\linewidth}{!}{\includegraphics{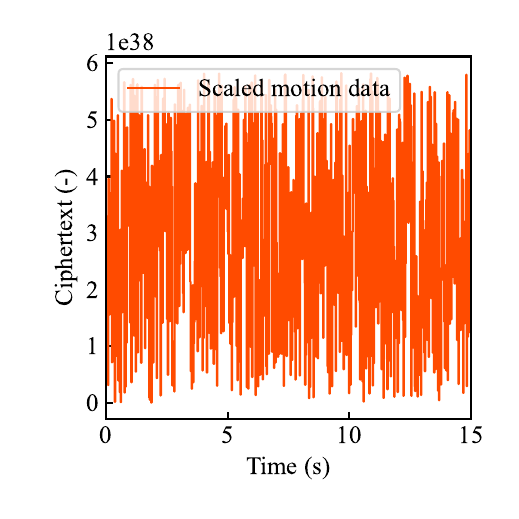}}}
\caption{The experimental results for object contact scenario; (a)-(d) saving motions; (e)-(h) loading motions} \label{ff}
\end{figure}

The object contact scenario represents saving and loading motion data involving physical object interaction, highlighting the capability of reproducing the contact force.
This experiment considers an aluminum block fixed on the basement as an obstacle within the movable range of the follower’s robotic arm, as shown in Fig.~\ref{obst}.
In saving motion, the follower robotic arm physically contacts the aluminum block in the telemanipulation by the operator, resulting in a 15-second motion sequence. 
In loading motion, the developed system reproduces the stored motion such that the follower emulates the leader’s movements during and after contact with the object, where the scaling parameter $\alpha$ was set to (1,1).
The obtained results are shown in Fig.~\ref{ff}.

Figs.~\ref{ff}(a)-(d) show the time responses of the leader and follower robotic arms during saving motion, where the red and blue lines indicate the leader and follower, respectively.
Figs.~\ref{ff}(a) and (b) illustrate the yaw-axis rotation angles and the estimated torque of the leader and follower, respectively.
Figs.~\ref{ff}(c) and (d) shows the encrypted motions regarding $\theta_l$ and $\hat{\tau}^{e}_l$ .
These figures confirm that the follower comes into contact with the aluminum object at $\SI{5}{s}$ while maintaining nearly identical rotation angles to the leader and that estimated torques for the leader and follower are equal magnitudes and opposite signs, affirming adherence to Newton’s third law. 
The encrypted rotation angles and estimated torques exhibiting random fluctuations at $10^{38}$ were stored in $\mathcal{D}$.

Figs.~\ref{ff}(e)-(h) show the time responses of the follower during loading motion, where the object is fixed on the same location as saving motion.
Figs.~\ref{ff}(e) and (f) illustrate the yaw-axis rotation angles and the estimated torques, respectively, where the meanings of the blue and red lines are the same as those in Fig.~\ref{pt}. 
Figs.~\ref{ff}(g) and (h) illustrate the leader’s encrypted motion loaded from $\mathcal{D}(\alpha)$.
Consequently, these figures confirm that even in the aluminum-block environment, the proposed system securely reproduces the follower’s behavior which was almost the same as the leader’s behavior.

\subsection{Spatial scaling}
\begin{figure}[t]
\centering
\subfloat[Rotation angle: $\theta_l$ and $\theta_f$.]{%
\resizebox*{.24\linewidth}{!}{\includegraphics{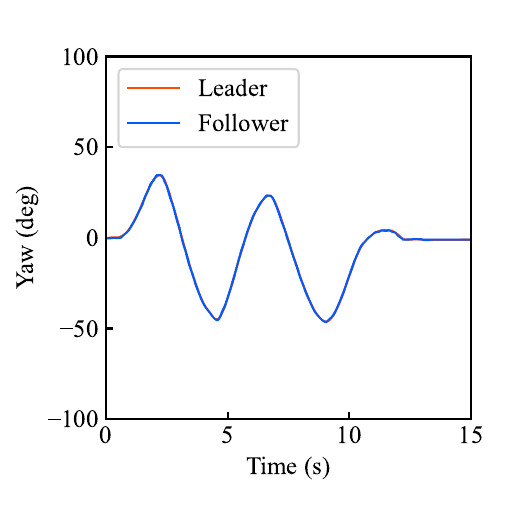}}}\ 
\subfloat[Estimated reaction force: $\hat{\tau}^{e}_l$ and $\hat{\tau}^{e}_f$.]{%
\resizebox*{.24\linewidth}{!}{\includegraphics{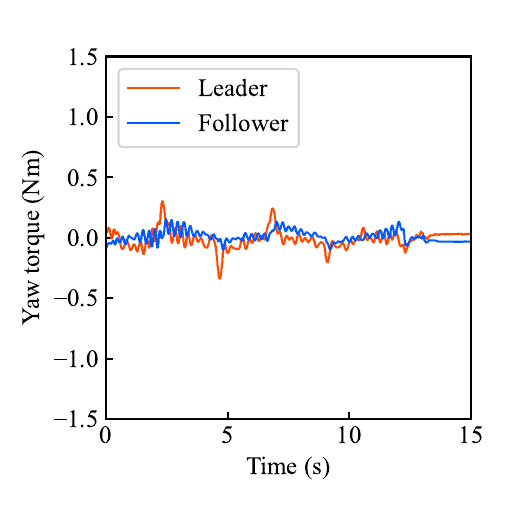}}}\ 
\subfloat[Encrypted rotation angle: $\Enc{(\theta_l)}$.]{%
\resizebox*{.24\linewidth}{!}{\includegraphics{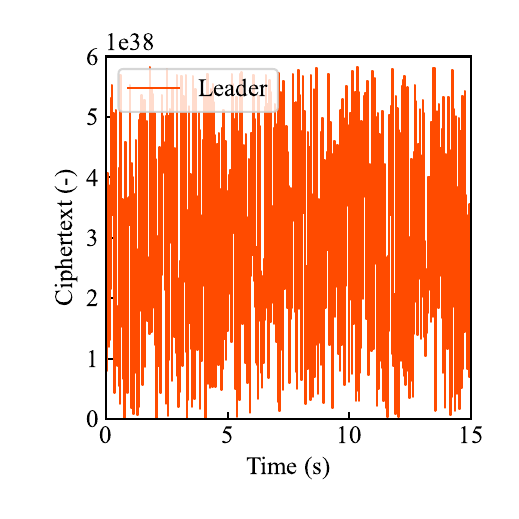}}}
\subfloat[Encrypted estimated torque: $\Enc{(\hat{\tau}^e_l)}$.]{%
\resizebox*{.24\linewidth}{!}{\includegraphics{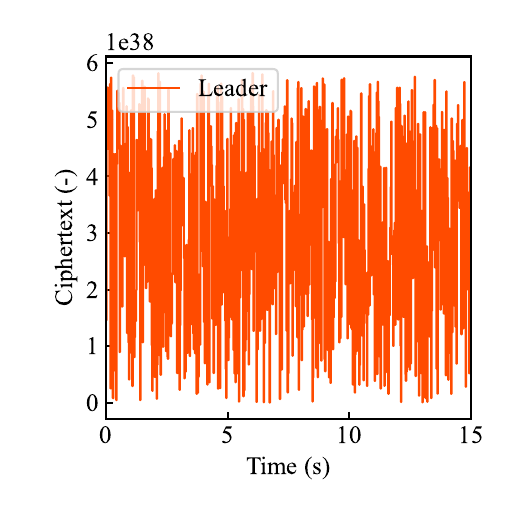}}}\\
\subfloat[Rotation angle: $\theta_l$ and $\theta_f$.]{%
\resizebox*{.24\linewidth}{!}{\includegraphics{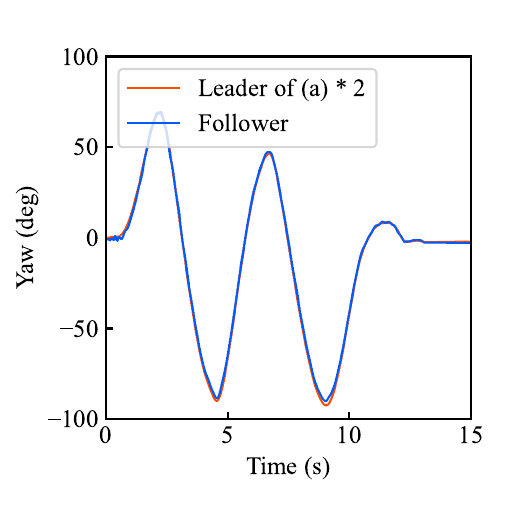}}}
\subfloat[Estimated reaction force: $\hat{\tau}^{e}_l$ and $\hat{\tau}^{e}_f$.]{%
\resizebox*{.24\linewidth}{!}{\includegraphics{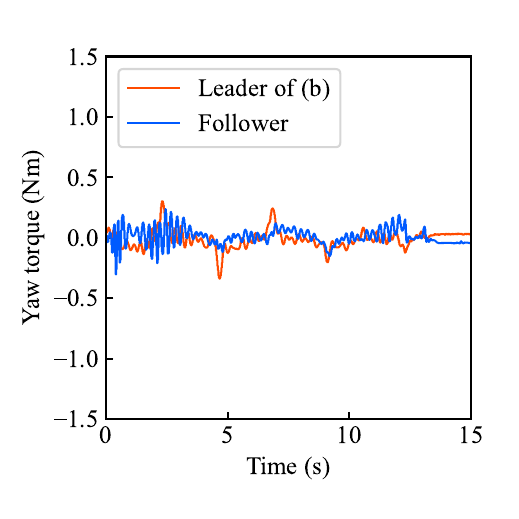}}} 
\subfloat[Encrypted rotation angle: $\Enc{(\theta_l)}$.]{%
\resizebox*{.24\linewidth}{!}{\includegraphics{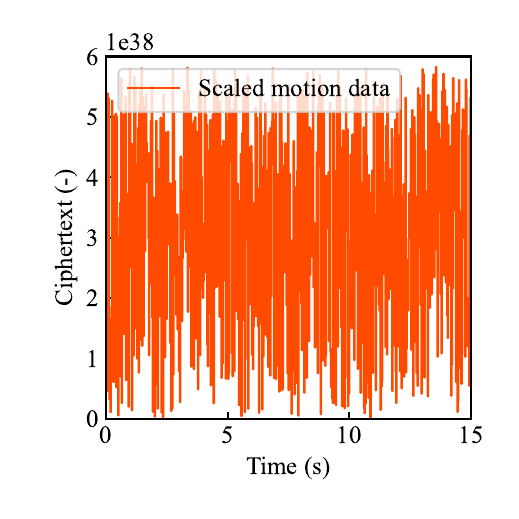}}}
\subfloat[Encrypted estimated torque: $\Enc{(\hat{\tau}^e_l)}$.]{%
\resizebox*{.24\linewidth}{!}{\includegraphics{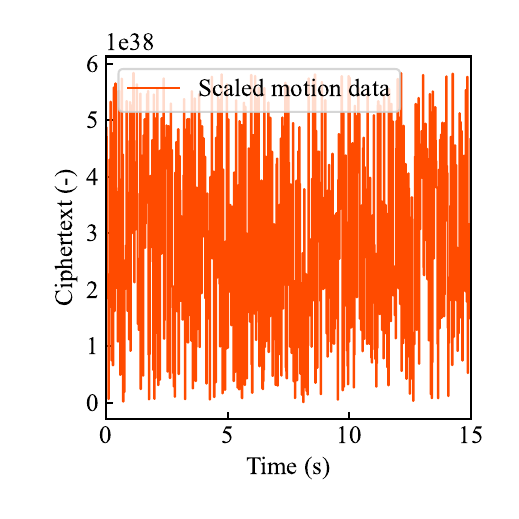}}}
\caption{The experimental results for free motion scenario with scaling parameter $\alpha=(2,1)$; (a)-(d) saving motions; (e)-(h) loading the scaled motions.} \label{em}
\end{figure}

\begin{figure}[t]
\centering
\subfloat[Rotation angle: $\theta_l$ and $\theta_f$.]{%
\resizebox*{.24\linewidth}{!}{\includegraphics{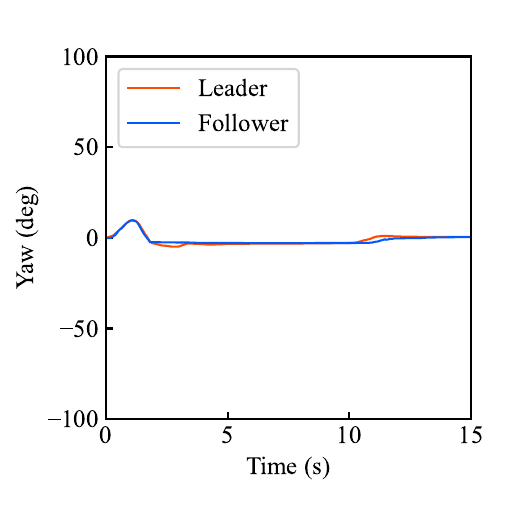}}}\ 
\subfloat[Estimated reaction force: $\hat{\tau}^{e}_l$ and $\hat{\tau}^{e}_f$.]{%
\resizebox*{.24\linewidth}{!}{\includegraphics{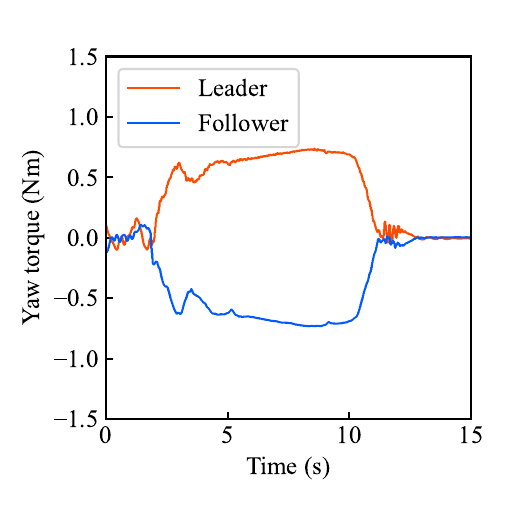}}}\ 
\subfloat[Encrypted rotation angle: $\Enc{(\theta_l)}$.]{%
\resizebox*{.24\linewidth}{!}{\includegraphics{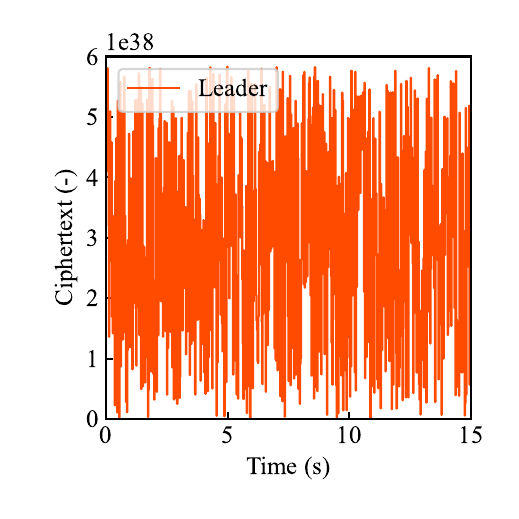}}}
\subfloat[Encrypted estimated torque: $\Enc{(\hat{\tau}^e_l)}$.]{%
\resizebox*{.24\linewidth}{!}{\includegraphics{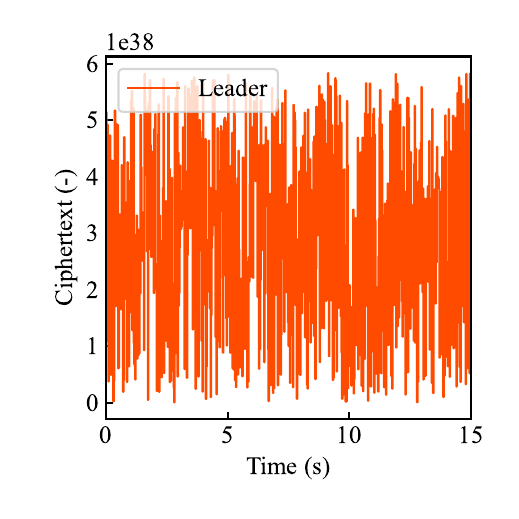}}}\\
\subfloat[Rotation angle: $\theta_l$ and $\theta_f$.]{%
\resizebox*{.24\linewidth}{!}{\includegraphics{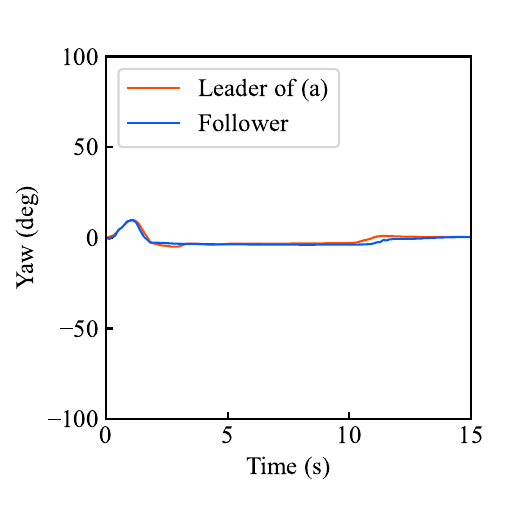}}}
\subfloat[Estimated reaction force: $\hat{\tau}^{e}_l$ and $\hat{\tau}^{e}_f$.]{%
\resizebox*{.24\linewidth}{!}{\includegraphics{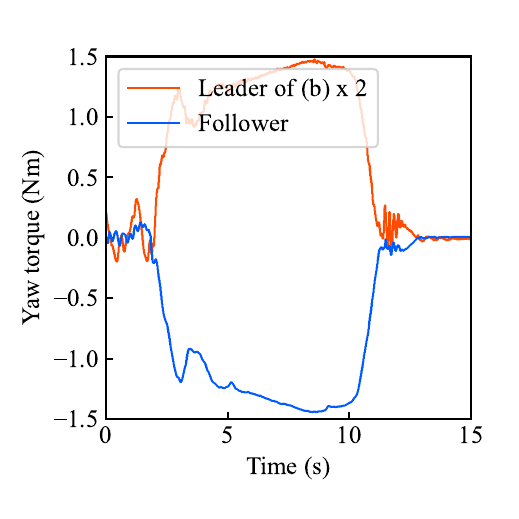}}} 
\subfloat[Encrypted rotation angle: $\Enc{(\theta_l)}$.]{%
\resizebox*{.24\linewidth}{!}{\includegraphics{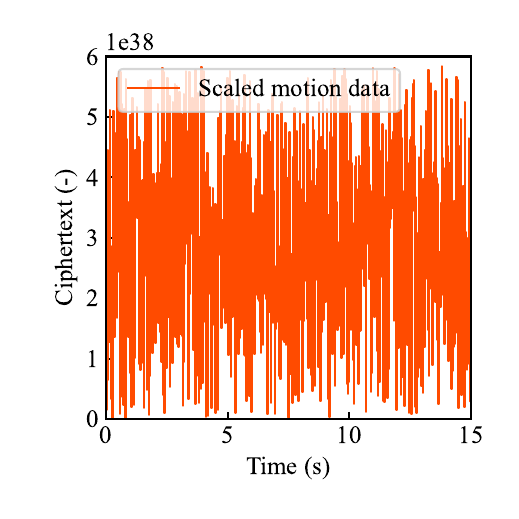}}}
\subfloat[Encrypted estimated torque: $\Enc{(\hat{\tau}^e_l)}$.]{%
\resizebox*{.24\linewidth}{!}{\includegraphics{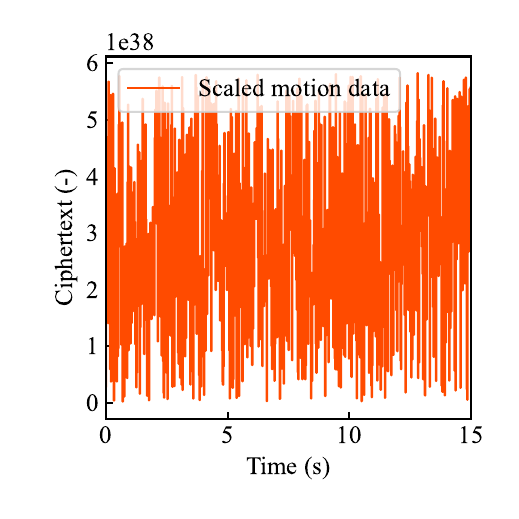}}}
\caption{The experimental results for object contact scenario with scaling parameter $\alpha=(1,2)$; (a)-(d) saving motions; (e)-(h) loading the scaled motions.} \label{ef}
\end{figure}

The spatial scaling scenario represents saving and loading the scaled motion with and without object contact.
This experiment investigates the system’s capability of reproducing the scaled motion in each scenario of free motion with $\alpha=(2,1)$ and object contact with $\alpha=(1,2)$. 
Similarly, in saving motion, 15-second motion sequences were stored in the secure memory $\mathcal{D}$, and in loading motion, the follower’s behavior was reproduced using the scaled motion loaded from $\mathcal{D}(\alpha)$.
The obtained results for the free motion and object contact scenarios are shown in Figs.~\ref{em} and \ref{ef}, respectively.

Figs.~\ref{em}(a)-(d) show the time responses of both the robotic arms during saving motion, which confirms that the follower’s behavior matches the leader’s behavior.
Meanwhile, Figs.~\ref{em}(e)-(h) show the time responses of the follower’s behavior during loading motion. 
Fig.~\ref{em}(e) confirms that the yaw-axis rotation angle of the follower was successfully doubled to the original leader’s behavior, while the estimated torque in Fig.~\ref{em}(f) has no significant impact.
Figs.~\ref{em}(g) and (h) illustrate the encrypted rotation angle and estimated torque that were loaded from $\mathcal{D}(\alpha)$.

Furthermore, Figs.~\ref{ef}(a)-(d) show the time responses of both the robotic arms during saving motion.
In this case, the follower comes into contact with the aluminum object at 2~s.
During contact, the leader maintains a rotation angle nearly identical to the follower’s, exhibiting an estimated torque of the opposite sign.
Meanwhile, Figs.~\ref{ef}(e)-(h) show the time responses of both the robotics arms during loading motion.
Fig.~\ref{ef}(f) confirms that the estimated torque of the follower was successfully doubled to the original leader's behavior, while the rotation angle in Fig.~\ref{ef}(e) has no significant impact.
Fig.~\ref{ef}(g) and (h) illustrate the scaled encrypted behavior that was loaded from $\mathcal{D}(\alpha)$.

\section{Conclusion}\label{sec:con}
This study developed the encrypted motion-copying system with the robotic arms in the encrypted control fashion to ensure secure motion preservation and reproduction.
The experimental validation across the three scenarios confirmed that the developed system could operate using the encrypted motions and scale the preserved motions for reproduction without decryption.

In future work, a fully homomorphic and keyed-homomorphic encryption scheme will be applied to construct motion-copying systems, bolstering the security level in managing preserved motions.
Moreover, we will perform the stability analysis of the developed control system, which overcomes the difficulty of addressing both nonlinearity and quantization errors.

\section*{Funding}
This work was supported by JSPS KAKENHI Grant Number 22H01509. 

\bibliographystyle{tfnlm}
\bibliography{bib/manuscript}

\end{document}